\documentclass[aps,prd,amsmath,amssymb,preprintnumbers,nofootinbib,a4paper]{revtex4}
\pdfoutput=1
\usepackage{graphicx}
\usepackage{color}
\newcommand{\beq}{\begin{eqnarray}}
\newcommand{\eeq}{\end{eqnarray}}

\newcommand{\bmp}{\noindent\begin{minipage}{16cm}}
\newcommand{\emp}{\end{minipage}\vskip 7mm} 

\usepackage{dcolumn}
\usepackage{bm}
\usepackage{bbm}
\usepackage{subfigure}
\usepackage{pxfonts}

\usepackage{epsfig}

\usepackage[ margin=5pt, font=normalsize,labelfont=bf,justification=raggedright]{caption}
\usepackage{youngtab}
\usepackage{slashed}

\definecolor{rossoCP3}{cmyk}{0,.88,.77,.40}

\def\lsim{\mathrel{\rlap{\lower4pt\hbox{\hskip1pt$\sim$}}
    \raise1pt\hbox{$<$}}}                
\def\gsim{\mathrel{\rlap{\lower4pt\hbox{\hskip1pt$\sim$}}
    \raise1pt\hbox{$>$}}}                

\begin{document}
\includegraphics[width=3.5cm]{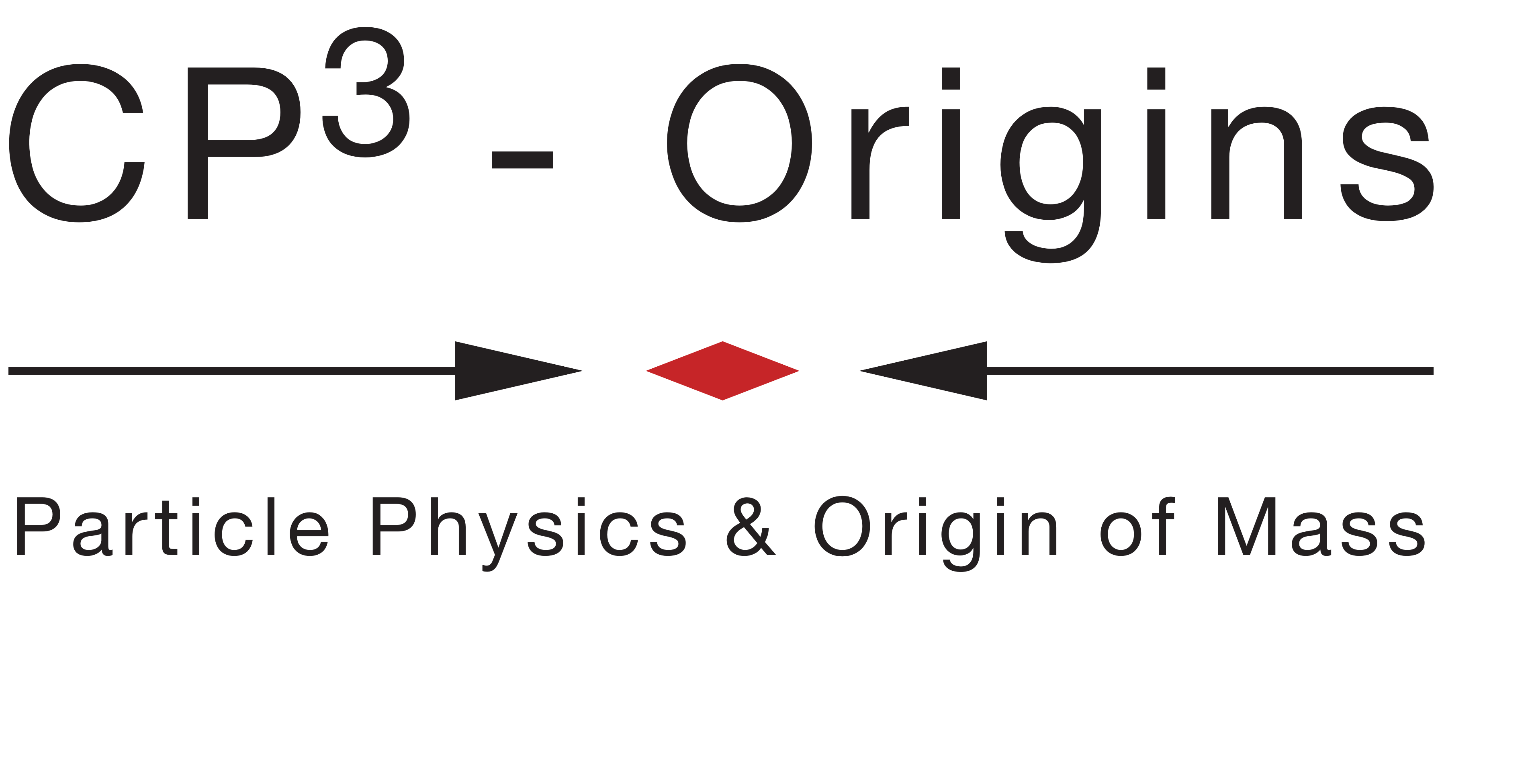}
\title{\Large  \color{rossoCP3}Composite Higgs to two Photons and Gluons  }
\author{Tuomas Hapola$^{\color{rossoCP3}{\varheartsuit}}$}\email{hapola@cp3-origins.net} 
\author{Francesco  Sannino$^{\color{rossoCP3}{\varheartsuit}}$}\email{sannino@cp3-origins.net} 
\affiliation{
$^{\color{rossoCP3}{\varheartsuit}}${ CP}$^{ \bf 3}${-Origins}, 
University of Southern Denmark, Campusvej 55, DK-5230 Odense M, Denmark.
}
\begin{abstract} 
We introduce a simple framework to estimate the composite Higgs boson coupling to two-photon in Technicolor extensions of the standard model.  The same framework allows us to predict the composite Higgs to two-gluon process. We compare the decay rates with the standard model ones and show that the corrections are typically of order one. We suggest, therefore, that the two-photon decay process can be efficiently used to disentangle a light composite Higgs from the standard model one.  We also show that the Tevatron results for the gluon-gluon fusion production of the Higgs either exclude  the techniquarks to carry color charges to the 95\% confidence level, if the composite Higgs is light, or that the latter must be heavier than around $200$~GeV.  
 \\
 [.1cm]
{\footnotesize  \it Preprint: CP$^3$-Origins-2011-04}
 \end{abstract}

\maketitle

\section{Introduction}

One of the main goals of the Large Hadron Collider experiments (LHC)  is to shed light on the mechanism responsible for the spontaneous breaking of the electroweak symmetry. 
 During the past decades many extensions to the standard model (SM) have been envisioned based on one or more principles or theoretical prejudices.  
 Arguably models of dynamical electroweak symmetry breaking, also known as Technicolor \cite{Weinberg:1979bn,Susskind:1978ms}, are among the most natural extensions since they replace the Higgs sector of the SM with a new strongly coupled four dimensional gauge theory which is free from fine tuning till the Planck scale. Understanding strong dynamics is therefore imperative when trying to build extensions of this type. New analytic methods are  currently being proposed in order to shed light on the nonperturbative aspects of the gauge theories relevant for Technicolor extensions of the SM. These new methods range from the use of dualities of  gauge-gauge type \cite{Sannino:2009qc,Sannino:2009me,Sannino:2010fh,Mojaza:2011rw,Terning:1997xy} to gauge-gravity ones \cite{Fabbrichesi:2008ga,Dietrich:2009af,Dietrich:2008ni,Dietrich:2008up,Nunez:2008wi} . The dynamics and spectrum of several gauge theories relevant for the construction of models of dynamical electroweak symmetry breaking are also being investigated using first principle lattice computations \cite{Catterall:2010du,Bursa:2009we,DeGrand:2010na,Hietanen:2009az,Hietanen:2008mr}.  A recent summary of these activities together with a basic introduction to Technicolor, its impact on cosmology, and the phase diagram of strongly coupled theories can be found in \cite{Sannino:2009za}. 

It has been argued in \cite{Hong:2004td,Sannino:2008ha}, using Large N arguments, and in \cite{Dietrich:2005jn,Dietrich:2006cm}, using the saturation of the trace of the energy momentum tensor,  that  models of dynamical electroweak symmetry breaking featuring (near) conformal dynamics  contain a composite Higgs state which is light with respect to the new strongly coupled scale ($4\,\pi \, v_{\rm EW}$ with $v_{\rm EW} \simeq 246$~GeV). These indications have led to the construction of models of Technicolor with a naturally {\it light composite} Higgs. Recent investigations using Dyson-Schwinger (SD) \cite{Doff:2008xx} and gauge-gravity dualities \cite{Fabbrichesi:2008ga} also arrived to the conclusion that the composite Higgs can be light \footnote{The Higgs boson here is identified with the lightest $0^{++}$ state of the theory. Calling it a dilaton or a meson makes no physical difference since these two states mix  at the 100\%  level and both couple to the trace of the stress energy momentum tensor of the theory. In the construction of the low energy effective theory saturating the trace anomaly there is no way to distinguish these states.}.  

The presence of a light composite Higgs immediately triggers the question: {\it How do we disentangle the SM Higgs from the composite one at the LHC?}. We have already partially answered this question in \cite{Belyaev:2008yj} by showing that the production of the Higgs in association with a SM gauge boson can be used to this goal. Another interesting work on this subject is \cite{Delgado:2010bb}.  Here we turn our attention to one of the most promising channels for the discovery of a light Higgs namely the Higgs to two-photon decay process.

In this work we introduce a simple framework to estimate the composite Higgs boson coupling to two photons by  using the saturation of the electromagnetic trace anomaly at the underlying gauge theory level.  We provide a general expression valid for technifermions transforming according to generic representations of the underlying Technicolor gauge group,  with or without carrying ordinary color charges. We then compare the decay rate with the SM one and show that the corrections are typically of order one, even if the techniquark degrees of freedom do not carry color. We suggest, therefore, that the two-photon decay process can be efficiently used to disentangle a light composite Higgs from a SM one. These large effects are due to the increase of the number of degrees of freedom with respect to the SM case  and therefore our estimates are   insensitive to the specific model computations. 

The formalism we introduce in order to estimate the coupling to two-photon allows us to readily estimate also the corrections to the composite Higgs to two-gluon process stemming from the Technicolor sector, if the techniquarks carry color charges. We will show that, in this case, the Tevatron results for the gluon-gluon fusion production of the Higgs sets very stringent constraints\footnote{A more careful analysis taking into account higher-order QCD corrections should be performed to provide a more precise constraint. However we expect an even more stringent constraint given that these corrections will typically enhance this process \cite{Spira:1995rr}.}.

We show that our simple approach can be used for any extension of the SM featuring a composite Higgs arising from natural four-dimensional strongly coupled dynamics. 

\section{Composite Higgs to gamma gamma}

The one loop Higgs decay width to two photons for any weakly interacting elementary particle contributing to this process can be neatly summarized via the formula \cite{Gunion:1989we}: 

\begin{equation}
\begin{split}
\Gamma(h\to\gamma\gamma)=\frac{\alpha^2G_{F}M_h^2}{128\sqrt{2}\pi^3}~\left\lvert \sum_i~n_i~Q_i^2~F_{i} \right\rvert^2,
\end{split}
\label{width}
\end{equation}
where i runs over the spins, $n_{i}$ is the multiplicity of each species with electric charge $Q_i$  in units of $e$. The $F_i$ functions are given by
\begin{equation}
F_1=2+3\tau+3\tau(2-\tau)f(\tau)\ ,\quad 
F_{1/2}=-2\tau[1+(1-\tau)f(\tau)]\ ,\quad 
F_0=\tau\left( 1-\tau f(\tau) \right)\ ,
\label{loopfun}
\end{equation}
where $\tau=\frac{4m_i^2}{M_h^2}$ and
\begin{equation}
  f(\tau) = \left\{ 
  \begin{array}{c c}
    \left( \arcsin \sqrt{\frac{1}{\tau}} \right)^{2},  & \quad \text{if $\tau ~\geq~1$}\\
    -\frac{1}{4}\left( \log \left( \frac{1+\sqrt{1-\tau} }{1-\sqrt{1-\tau}} \right) -i\pi \right)^2,  & \quad \text{if $\tau~<~1$}\\
  \end{array} \right. 
\end{equation}
The lower index of the function $F$ indicates the spin of each particle contributing to the process. It is clear from this formula that there can be strong interferences between the different terms contributing. In particular we anticipate that there is an important interplay between the SM gauge bosons and the Technicolor contribution for this process. 

The contribution to the Higgs to two-photon process in the SM is due to the $W$s and SM fermion loops (mostly the top) contributions. In fact it is an excellent approximation to use just the top in the fermion loop, since the ratio between the result using only the top and the full one is around $0.98$ for the reference value of the Higgs mass of $m_h \simeq 120$GeV. The approximation improves as the Higgs mass increases.  
We consider here the case in which the Higgs is a composite state constituted of some more fundamental matter as in Technicolor. The SM Higgs sector Lagrangian is therefore interpreted simply as a low energy effective theory. We assume the underlying Technicolor theory to have global symmetry $SU_L(2)\times SU_R(2)$ and therefore the composite Higgs effective field coincide with the SM one: 
\begin{equation}
H = \frac{1}{\sqrt{2}} \left(
\begin{array}{c}
  \pi_2 + i \pi _1    \\
  \sigma  - i \pi_3
\end{array}
\right) \ .
\end{equation} 
We arrange the Technicolor dynamics in such a way that the scalar field $\sigma$ acquires the vacuum expecation value  $\langle \sigma \rangle = v_{EW}$ yielding the gauge boson masses  $M_W^2  = g^2 v^2_{EW} /4$. It is straightforward to generalize the global symmetries of the Technicolor theories to large symmetry groups.

The contribution to the sought process is modeled by re-coupling, in a minimal way, the composite Higgs to the techniquarks $Q$ via the following operator: 
\begin{equation} 
L_{QH} = \sqrt{2}\, \frac{M_Q}{v_{EW}} \left[ {\overline{Q}_L}^t\cdot H {D_R}_t + {\overline{Q}_L}^t\cdot (i\, \tau_2 \, H^{\ast}) {U_R}_t    \right] + {\rm h.c.} \ ,  
\end{equation}
where $M_{Q}$ is the dynamical mass of the techniquark and $t = 1,\dots, d[r]$ is the Technicolor index and $d[r]$ the dimension of the representation under which the techniquarks transform. This type of hybrid models have been employed many times in the QCD literature to extra dynamical information and predictions for different  phenomenologically relevant processes. {}A close example is  the $\sigma(600)$ decay into two photons \cite{Pennington:2006dg} investigated using, for example, hybrid type models in  \cite{Giacosa:2007bs} and \cite{vanBeveren:2008st}. 

We assumed a single Technicolor doublet with respect to the Weak interactions as predicted by minimal models of Technicolor \cite{Sannino:2004qp}: 
\begin{equation}
Q_L^t =
\left(
\begin{array}{c}
  U^t   \\
  D^t    
\end{array} 
\right)_L  \ , \qquad U_R^t \ , \quad D_R^t \ .
\end{equation}
The following generic hypercharge assignment is free from gauge anomalies:
\begin{align}
Y(Q_L)=&\frac{y}{2} \ ,&\qquad Y(U_R,D_R)&=\left(\frac{y+1}{2},\frac{y-1}{2}\right) \ ,  \nonumber\\
Y(L_L)=& -3\frac{y}{2} \ ,&\qquad
Y(N_R,E_R)&=\left(\frac{-3y+1}{2},\frac{-3y-1}{2}\right) \ .
 \label{assign2}
\end{align}
We added a new Lepton family to cure the Witten topological anomaly (with respect to the weak interactions) \cite{Witten:1982fp} when the Technicolor sector features an odd number of doublets charged under the $SU(2)_L$ Weak interactions. This occurs when the dimension of the Technicolor matter representation is an odd number. A further Leptonic interaction term is then added: 
\begin{equation} 
L_{LH} = \sqrt{2}\, \frac{M_E}{v_{EW}} \, {\overline{L}_L}\cdot H {E_R}+  \sqrt{2}\, \frac{M_N}{v_{EW}}{\overline{L}_L}\cdot (i\, \tau_2 \, H^{\ast}) {N_R}    + {\rm h.c.} \ ,
\end{equation}
with $M_E$ and $M_N$ the fermion masses expected to be of the order of the electroweak scale \cite{Frandsen:2009fs}. The detailed spectrum is partially dictated by the precision electroweak constraints.  

The techniquark dynamical mass is intrinsically linked to the technipion decay constant $F_{\pi}$ via the Pagels-Stokar formula  
$M_Q  \approx \frac{2\pi F_{\pi}}{\sqrt{d(r)}}$. 
  To achieve the correct values of the weak gauge bosons  the decay constant is related to the electroweak scale via: 
\begin{equation}
v_{\text{EW}}=\sqrt{N_{f}}F_{\pi}, \qquad \text{with}\qquad v_{\text{EW}}=246~ \text{GeV} \ .
\end{equation}  
 We have checked that  the numerical value of  the contribution of the $M_{E/N}$ states is negligible when the masses are of the oder of the electroweak scale. We have used as a reference value for these lepton masses  $500~\text{GeV}$ \footnote{For  $M_{E/N}>>M_{h}$ the contribution approaches a constant due to the functional form of $F_{1/2}$.}. 
 
 At this point we determine the contribution of the techniquarks, not carrying ordinary color charge, and new leptons for the diphoton process by evaluating the naive one loop triangle diagrams in which the composite Higgs to two techniquarks, as well as the new leptons, vertex can be read off from the Lagrangian above. The contribution at the amplitude level for each new particle is identical to the one of a massive SM fermion provided one takes into account the correct electric charge and degeneracy factor. 

Although we adopted a simple model computation we now argue that the final geometric dependence on the gauge structure, matter representation and number of techniflavors of the composite theory is quite general.

\begin{figure}[htbp]
\begin{center}
 \includegraphics[width=0.5\textwidth]{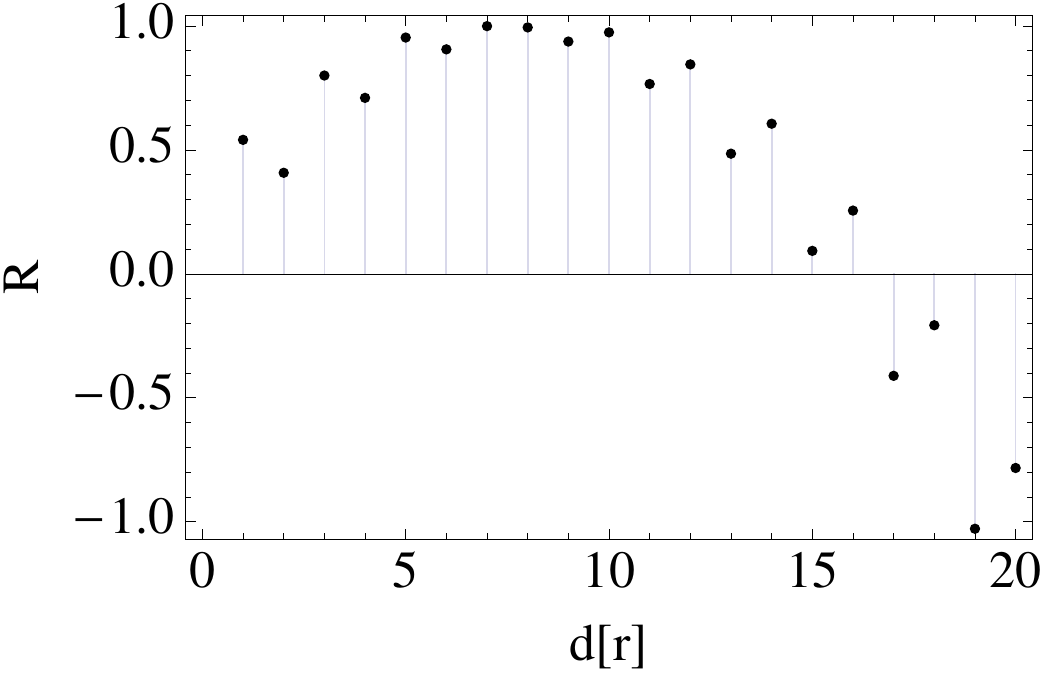}
\caption{Difference of the Walking TC and SM decay widths divided by the SM width plotted with respect to number of colors. $m_{h}=120$.}
\label{color}
\end{center}
\end{figure}

\begin{figure}[htbp]
\begin{center}
 \includegraphics[width=0.5\textwidth]{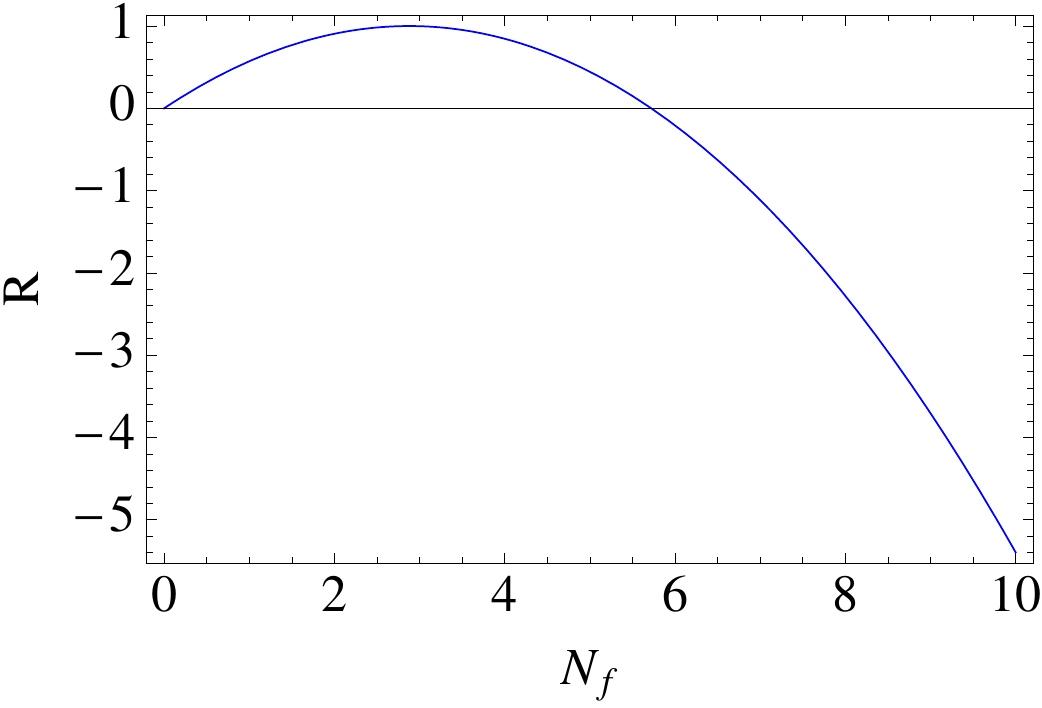}
\caption{Difference of the Walking TC and SM decay widths divided by the SM width plotted with respect to number of flavors. $m_{h}=120$.}
\label{flavor}
\end{center}
\end{figure}

\begin{figure}[htbp]
\begin{center}
 \includegraphics[width=0.5\textwidth]{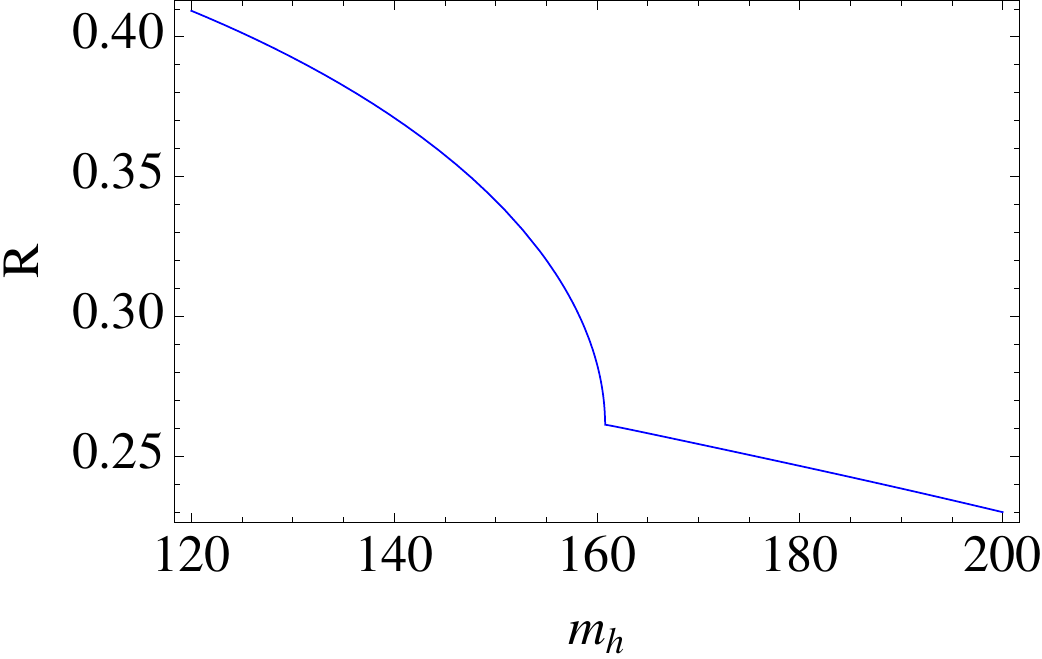}
\caption{ Dependence of $R$ as function of the Higgs mass $m_{h}$ for a reference value of the techniquark matter representation $d[r] =2$ and two number of Dirac techniflavors.}
\label{Higgs-dependence}
\end{center}
\end{figure}

We plot in Fig.~\ref{color} the intrinsic dependence on the dimension of the Technicolor matter representation $d[r]$ raccording to the ratio:
\begin{equation}
R=\frac{\Gamma_{SM}(h\to \gamma\gamma)-\Gamma_{TC}(h\to \gamma\gamma)}{\Gamma_{SM}(h\to \gamma\gamma)}.
\label{haa-ratio}
\end{equation}
 For any odd representation we included the Lepton contribution and therefore we could not simply join the points. Interestingly when $d[r]$ is around $7$  the contribution from the techniquarks adds to one of the SM fermion ones (i.e. mainly the top) and cancels the one due to the W's  leading to $\Gamma_{TC}\approx 0$.  The techniquark matter dominates the process for $d[r]$ larger than around seven.  A similar behavior occurs when we increase the number of flavors as shown in Fig. \ref{flavor}.   
The dependence on the Higgs mass with $d[r]=2$ and $N_f=2$ is shown in Fig.~\ref{Higgs-dependence}.
The $WW$ threshold is evident in the plot which shows how the ratio decreases with the Higgs mass. 
  
\begin{figure}[htbp]
\begin{center}
 \includegraphics[width=0.5\textwidth]{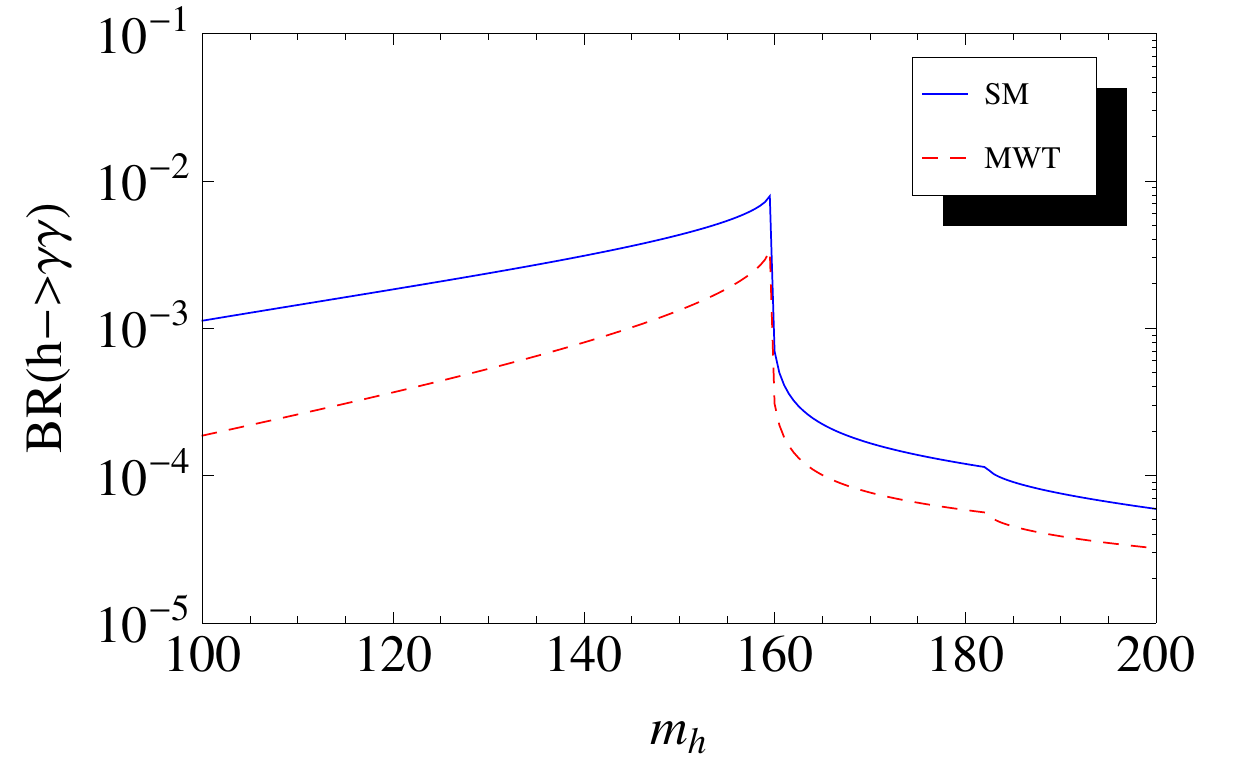}
\caption{ Higgs to two photons branching ratio in Minimal Walking Technicolor as function of the composite Higgs mass versus the SM result.}
\label{H2Gbranching}
\end{center}
\end{figure}

 To illustrate our results we plot in Fig.~\ref{H2Gbranching} the Higgs to two-photon branching ratio as function of the Higgs mass for the Minimal Walking Technicolor model and compare it to the SM one.   
It is clear from the plot that  we expect sizable corrections to the SM Higgs to two photons process to be visible at the LHC.

 \section{Composite Higgs to Gluon-Gluon Processes}
 
 It is interesting to estimate the impact on the digluon process coming from a potentially composite Higgs made by technimatter charged under QCD interactions. This contribution can be estimated, as we have done for the diphoton process, by considering the one loop contribution of the techniquarks arriving at: 
\begin{equation}
\begin{split}
\Gamma(h\to gg)=\frac{\alpha_{s}^2G_{F}M_h^3}{64\sqrt{2}\pi^3}~\left\lvert \sum_i~F_{i}^{\rm SM}+\sum_j~2T[\tilde{r}_{j}]d[r_{j}]F_{j}^{\rm TC} \right\rvert^2,
\end{split}
\end{equation}
where $T[\tilde{r}_{j}]$ is the index of the representation of the techniquarks under the SM color group $SU(3)$ and $d[r_{j}]$ is the dimension of  the representation under the technicolor gauge group.
\begin{figure}[htbp]
\begin{center}
 \includegraphics[width=0.4\textwidth]{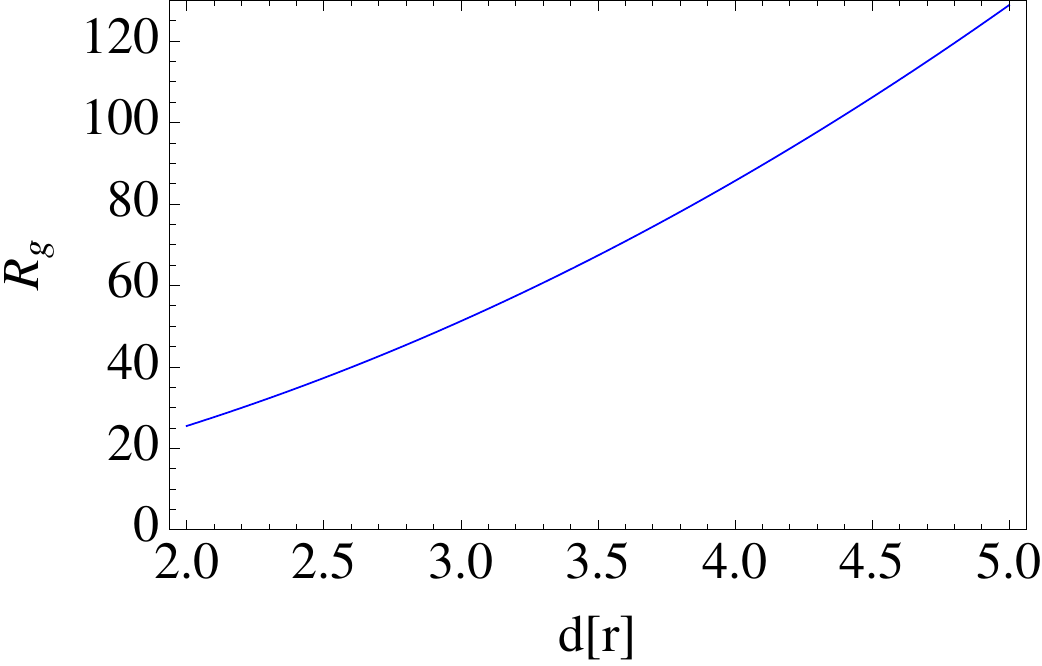}
\caption{Difference of the Walking TC and SM decay widths divided by the SM width plotted with respect to number of flavors. $m_{h}=120$.}
\label{ggh-mh}
\end{center}
\end{figure}
{}For  techniquarks transforming according to the fundamental representation of QCD the two gluon decay width of the Higgs is strongly enhanced. This is presented in Fig. \ref{ggh-mh} where we have plotted the ratio
 \begin{equation}
R_{g}=\frac{\Gamma_{TC}(h\to gg)-\Gamma_{SM}(h\to gg)}{\Gamma_{SM}(h\to gg)}.
\end{equation}
as a function of the dimension of the representation under the Technicolor gauge group. The composite higgs to two-gluon decay width increases monotonically with respect to the number of Technicolors in contrast with the diphoton decay width. The difference resides in the fact that gluons do not couple directly to the other gauge bosons of the SM.

We  now estimate the viability of the models with colored techniquarks by comparing our results with the study performed in \cite{Aaltonen:2010sv}.  Here the authors considered the effects of the fourth fermion SM generation to the process  $\sigma_{gg\to h}\times \text{BR}(h\to W^{+}W^{-})$. The fourth generation quarks enhances the $gg\to h$ production cross section by a factor  of $9$ to $7.5$ in the Higgs mass range $m_{h}=110-300$ GeV with respect to the SM result. This enhancement leads to a 95\% confidence level exclusion of the Higgs in the mass range $m_{h}=131-208$ GeV or of the presence of a fourth generation if the Higgs turns out to be within this mass energy range.  Colored techniquakrs are further penalized with respect to the SM fourth generation of quarks since they carry also technicolor leading to a further dramatic enhancement of this process.  \begin{figure}
\begin{center}
 \includegraphics[width=0.35\textwidth]{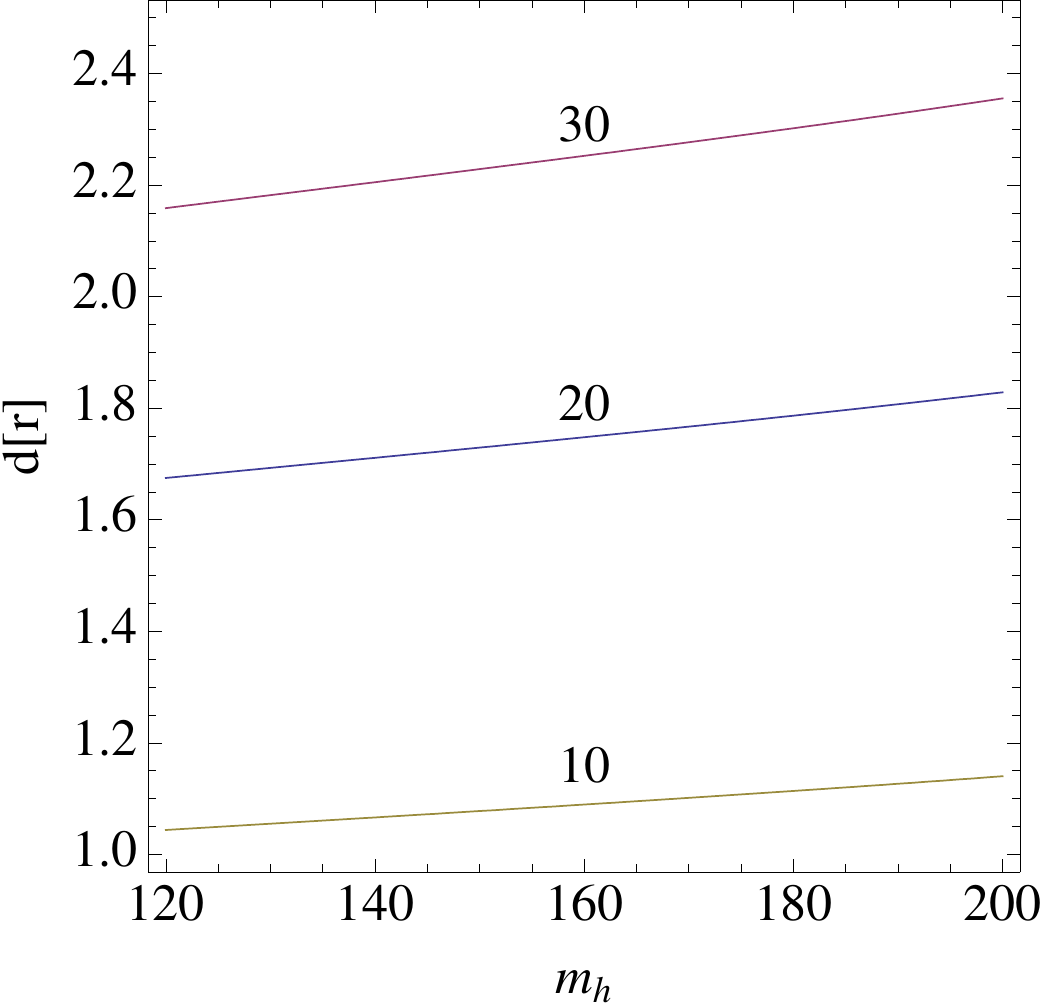}
\caption{Contours in the $d[r]$ versus Higgs mass plane for fixed values of the enhancement of the $gg\to h$ production cross section compared to the SM one. Obviously, in practice,  the dimension of the representation cannot be taken to be continuous and the plot simply shows that already for $d[r]$ around two (which is the minimal nontrivial matter group-theoretical dimension) the enhancement factor is around 30.  }
\label{contour}
\end{center}
\end{figure}
In Fig. \ref{contour} we have plotted contours in the dimension of the techniquark representation versus the Higgs mass when the $gg\to h$ production cross section is enhanced by factors 10, 20 and 30 respectively compared to the SM one. Obviously, in practice,  the dimension of the representation cannot be taken to be continuous and the plot simply shows that already for $d[r]$ around two (which is the minimal nontrivial matter group-theoretical dimension) the enhancement factor is around 30.  Thus we conclude that this process alone strongly indicates that viable Technicolor models should feature {\it colorless} techniquarks as it is the case of Minimal Walking Technicolor or that a light composite Higgs made of colored techniquarks is excluded by the Tevatron experiment. These results are in perfect agreement with the LEP electroweak precision data which require a small $S$ parameter.

\section{Conclusions} 
We introduced a simple framework to estimate the composite Higgs boson coupling to two photons and gluons for generic  Technicolor extensions of the SM.  By comparing the decay rates with the SM ones for the diphoton case we show that the corrections are typically of order one and therefore this process can be efficiently used to disentangle a light composite Higgs from a SM one.  We then turned our attention to the composite Higgs to two gluon process and showed that the Tevatron results for the gluon-gluon fusion production of the Higgs exclude either the techniquarks to carry color charges up to the 95\% confidence level or the existence of a light composite Higgs made of colored techniquarks.

\acknowledgments
We thank J.R. Andersen, G. Azuelos and M. T. Frandsen for a careful reading of the manuscript, relevant comments and suggestions.


\end{document}